\documentclass[reprint,amssymb, amsmath, aip,cha,twocolumn]{revtex4-1}
\usepackage{graphicx}
\usepackage[caption = false]{subfig}
\usepackage{color}
\usepackage{epsfig}
\usepackage{ifpdf}
\usepackage{url}%
\usepackage{cleveref}
\usepackage{bm}
\usepackage[colorlinks=true,linkcolor=blue]{hyperref}%
\expandafter\ifx\csname package@font\endcsname\relax\else
 \expandafter\expandafter
 \expandafter\usepackage
 \expandafter\expandafter
 \expandafter{\csname package@font\endcsname}%
\fi
\begin{document}
\title{Experimental Observation of a Dusty Plasma Crystal in the Cathode Sheath of a DC Glow Discharge Plasma}%
\author{M. G. Hariprasad}%
\email{hari.prasad@ipr.res.in}
\author{P. Bandyopadhyay}
\author{ Garima Arora}
\author{A. Sen}
\affiliation{$^{1}$Institute For Plasma Research, HBNI, Bhat, Gandhinagar, Gujarat, India, 382428}%
\date{\today}
\begin{abstract}
Dusty plasma crystals have traditionally been observed and studied in radio frequency (RF) discharge plasmas and their formation in a DC glow discharge plasma remains experimentally challenging. We report the first ever observation of a stable dusty plasma Coulomb crystal in the cathode sheath region of a DC glow discharge plasma. The observations are made in the DPEx device where crystals of mono-disperse Melamine Formaldehyde grains are produced in the background of an Argon plasma. The crystalline nature of the structure is confirmed through a host of measurements that includes the radial pair correlation function, Voronoi diagram, Delaunay Triangulation, the structural order parameter, the dust temperature and the Coulomb coupling parameter. The special features of the DPEx device that permit such a crystal formation are delineated and some principal physical features of the crystal discussed. 
\end{abstract}
\maketitle
\section{Introduction}\label{sec:intro}
A dusty plasma comprises of electrons, ions, neutrals and micron sized charged dust particles. The great diversity in the space and time scales of these constituent components make for a rich collective dynamics of this medium and has made dusty plasmas an active field of research for the last three decades or so \cite{pkshukla,goreerfanddc,morfill2009complex,firstcrystal}. One of the spectacular phenomena that can occur in a dusty plasma is the formation of an ordered arrangement of the dust component with characteristic features of a crystalline structure. This happens because the dust component can develop strong correlations due to the large charge on each dust particle and its low thermal energy. When the Coulomb parameter $\Gamma$, quantifying the ratio of the dust electrostatic potential energy to its thermal energy, exceeds a critical value the system can undergo a phase transition from a liquid state to a solid (crystalline) phase. The first experimental observation of such a Couloumb dusty plasma crystal, by two independent groups \cite{LinI94,firstcrystal} in 1994, opened up a whole new area of research in dusty plasma physics. 
 Dusty plasma crystals provide an excellent experimental platform for investigating a host of fundamental physics problems associated with phase transitions and allied topics that have relevance for areas as diverse as statistical mechanics, soft condensed matter, strongly and weakly coupled systems, active matter dynamics and warm dense matter. The ease of observation of the individual dust dynamics coupled with the convenient time scales of the collective dynamics of such systems have spurred a great deal of laboratory investigations of such crystals. Some of these investigations include melting processes in two and three dimensional crystals \cite{plasmacrystalmelting,fullmelting, ShockMelting}, dust lattice waves \cite{wakes,dustlatticewaves}, heat transport \cite{HeatTransfer,heattransport}, viscosity \cite{Viscosity}, recrystallization\cite{Recrystallization}, instabilities \cite{modeinstability}, dust oscillations in magnetic field\cite{magneticfield}, effect of magnetic field on phase transition \cite{magneticfieldphasetransition}, photophoretic force \cite{Photophoretic} and entropy production \cite{naturepaper}.  \par 

Most of the above laboratory investigations have been carried out on dusty plasma crystals created in a RF discharge plasma \cite{firstcrystal,Dustyplasmacavities,modeinstability,PhaseSeparation,Photophoretic,naturepaper,wakes,
dustlatticewaves} and there are very few reported studies of dusty plasma crystals in DC glow discharge plasmas.  Among such few works are those of Vladimir \textit{et al.} \cite{dcglowdischarge} and Maiorov \textit{et al.} \cite{dcgasmixture} who reported the formation of such crystals by the trapping of dust particles in the standing striations that appear under certain conditions in the positive column of a DC glow discharge.  One of the limitations of investigating crystals created in such a manner is that the strong variations in the electric field in the striations lead to small scale inhomogeneities in the particle clouds. Another disadvantage is the narrow range of discharge parameters over which such crystals can be formed and sustained. This prevents the exploration of such structures over a wide parametric domain. Mitic {\it et al} \cite{Mitic2008} were successful in avoiding the formation of striations by adjusting the parameters of the plasma discharge and trapping the dust particles in the plasma sheath region. They used a very high discharge voltage ($\sim 1000\;V$) which led to the creation of a longitudinal electric field and a concommitant large directed ion flow. To mitigate the effects of the longitudinal electric field on the particles in the plasma, they switched the polarity of the voltage between the electrodes with a frequency of 1 kHz thereby substantially reducing the directed flow of ions in a time averaged manner. While the detailed underlying reasons behind the difficulty of forming dust crystals in a DC glow discharge, compared to a RF discharge, are not fully understood there appear to be two contributory factors. These are lower charging of the dust particles and excessive heating of the dust by ion bombardment.  Since in a DC plasma the instantaneous and time-averaged electric fields are the same \cite{goreerfanddc} the dust and the electrons behave similarly despite their tremendously different masses.
In RF plasmas, on the other hand, electrons can respond to megahertz frequency reversals of the electric field, whereas the heavier dust particles cannot. As a result, the electrons can easily move into spatial regions where dust cannot. This difference between DC and RF plasmas has a significant effect on the charging of dust particles \cite{goreerfanddc}. In a DC plasma, the sheath has an insignificant electron density and therefore a dust particle that is immersed deep inside a DC sheath cannot collect many electrons. So the charge acquired by a dust particle in the DC sheath region is considerably less as compared to one in a RF sheath and consequently the Coulomb parameter (for the same dust temperature) tends to be relatively lower. The second factor, namely the heating due to the impact of streaming ions on the dust particles is related to the geometry of the conventional DC glow discharge systems where the electrodes are symmetric and placed directly facing each other. The ions accelerated by the DC voltage then directly impinge on the levitated dust cloud that is formed between the electrodes and causes its temperature to rise. This lowers the Coulomb parameter thereby inhibiting the formation of a crystalline structure. Any provision to mitigate these constraints in a conventional DC glow discharge set up could facilitate the formation of dust crystals in such devices. We believe our present system configuration succeeds in doing that as will be discussed later. \par

In this paper we report the first observation of a dusty plasma Coulomb crystal in the cathode sheath region of a DC glow discharge plasma. The experiments have been carried out in the DPEx device \cite{jaiswal2015dusty} with a unique electrode configuration that permits investigation of dust Coulomb crystals over a wide range of discharge parameters with excellent particle confinement. The crystalline character of the configurations have been confirmed by a variety of diagnostic analysis including the pair correlation function, the Voronoi diagram, the structural order parameter, the dust temperature and the coupling parameter that are obtained from the time evolution images of the particle positions.  \par
The paper is organized  as follows. Section II contains a brief description of the experimental set-up (the DPEx device) and the configuration of the electrodes. Section III presents our main results of the crystal formation along with its analysis using various methods. Section IV provides some concluding remarks including a discussion on the possible advantages arising from the present configuration that facilitates crystal formation. \\
\section{Experimental Set-up}\label{sec:setup}
\begin{figure}[ht]
\includegraphics[scale=1.0]{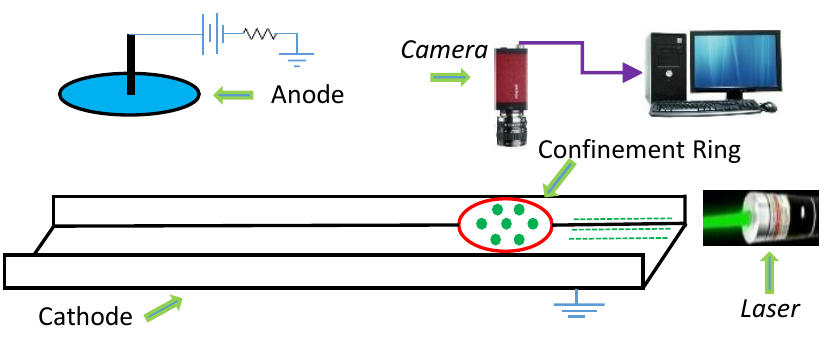}
\caption{\label{fig:fig1} A schematic diagram of dusty plasma experimental (DPEx) setup.  }
\end{figure}

Our experiments were carried out in the DPEx device, consisting of a $\Pi$-shaped glass vacuum chamber in which an Argon plasma is created by striking a discharge between a disc shaped steel anode and a long grounded steel plate as a cathode. Fig.~\ref{fig:fig1} provides a schematic diagram of the experimental arrangement. More details about the DPEx device and its operational characteristics are given in Jaiswal {\it et al} \cite{jaiswal2015dusty}.
A base pressure of $0.1$~Pa is achieved in the chamber with a rotary pump and the working pressure is set to 8-12 Pa by adjusting the pumping rate and the gas flow rate. Argon plasma is then produced by applying a discharge voltage of 280-320 DC volts between the anode and the cathode, and the discharge current is measured to be 1-4 mA. Mono-dispersive melamine fluoride (MF) particles of size 4.38 $\mu$m are then introduced in the chamber by a dispenser to form the dust component. These dust particles get negatively charged by accumulating more electrons (due to their higher mobility) than ions and levitate in the cathode sheath region due to a balance between the gravitational force and the vertical electrostatic force of the sheath of a metal ring of diameter 5 cm placed on the tray as shown in  Fig.~\ref{fig:fig1}. The ring facilitates the radial confinement and also gives the crystals a circular shape.  The dust particles usually levitate at the centre of the ring at a height of $\sim$ 1-2 cm above the cathode plate such that the laser light gets scattered from the particles. In addition, it is possible to vary the discharge parameters to manipulate the sheath around the ring and cathode to control the crystal size and  the crystal levitation height, respectively. The confinement ring also influences the dynamics of the streaming ions in a significant manner as will be discussed later. A green laser is used to illuminate the micron sized particles and the Mie-scattered light is captured by a CCD camera and stored into a computer for further analysis. \par

\section{Experimental Results and Data Analysis}
\subsection{Crystal structure}
\begin{figure}[ht]
\includegraphics[scale=1.0]{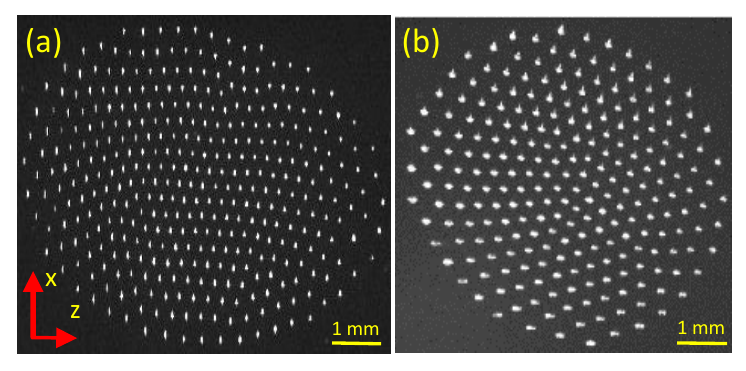}
\caption{\label{fig:fig2} Direct camera images of dusty plasma crystals obtained for dust particle sizes of (a)  $4.38 \mu m$ and b) 10.66 $\mu$m.}
\end{figure}

A crystalline structure of the dust mono-layer is obtained by a careful manipulation of the neutral pressure for a given discharge voltage. Fig.~\ref{fig:fig2}.(a) shows a camera image of a typical Coulomb coupled dusty plasma crystal with dust particles of diameter 4.38 $\mu m$ at a discharge voltage of 300 V and a neutral pressure 10 Pa. The crystal is circular in shape due to the horizontal (radial) confinement provided by the circular metal ring.  The structure is seen to be made up of hexagonal cells which display a good translational periodicity indicative of long range order. 
We have been successful in obtaining such crystalline structures over a range of neutral pressures (8 Pa to 12 Pa), different discharge voltages (280 V to 360 V) and for two different particle sizes, namely of diameter 4.38 $\mu m$ as shown in Fig.~\ref{fig:fig2}.(a)  and diameter 10.66 $\mu$m as shown in Fig.~\ref{fig:fig2}.(b). Once formed these crystals can be stably maintained for hours and can be conveniently used to conduct a variety of detailed parametric studies. To confirm and establish the true crystalline nature of the structure we have also carried out a set of diagnostic tests on the visual data in the form of calculating the pair correlation function, constructing a Voronoi diagram, estimating the dust temperature and the Coulomb coupling parameter. Below we provide a detailed description of the results of our analysis. 
\subsection{Radial Pair Correlation function}
\begin{figure}[ht]
\includegraphics[scale=0.75]{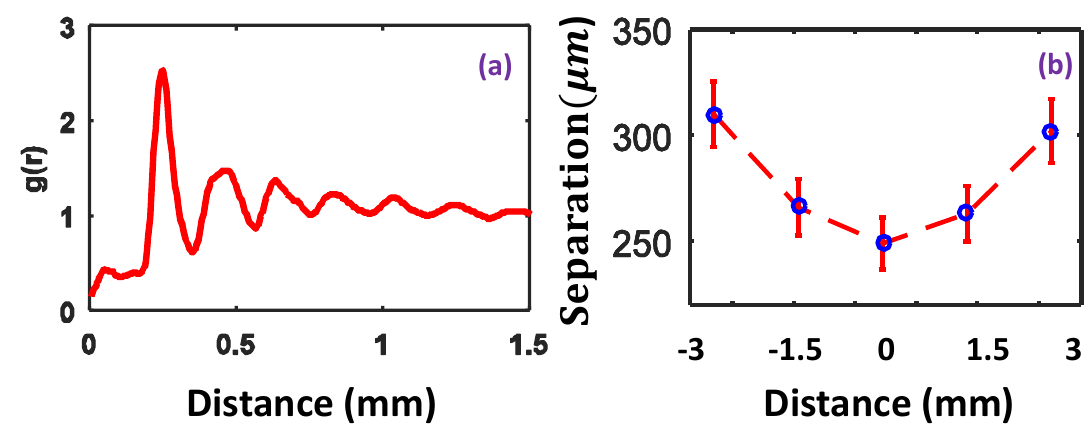}
\caption{\label{fig:fig3} (a) Pair correlation function corresponds to the crystal shown in Fig.~\ref{fig:fig2}.(a)  (b) Spatial variation of inter-particle distance.}
\end{figure}

The radial pair correlation function (RPDF) \cite{paircorrelation}, $g(r)$, is a measure of the probability of finding other particles in the vicinity of a given particle and provides useful information about the structural properties, such as the range of order, of a system. The relevant pair correlation function for the structure in Fig.~\ref{fig:fig2}.(a), calculated by taking into account all pairs of particles, is plotted in  Fig.~\ref{fig:fig3}.(a). The occurrence of multiple periodic peaks in this function establishes the long range ordering of the structure and indicates the formation of a crystalline structure. The position of the first peak provides information on the inter-particle distance and its variation with discharge parameters can serve as a useful diagnostic to study the spatial distribution of the particles in the crystal.  For the crystal shown in Fig.~\ref{fig:fig2}.(a) the average inter-particle distance is $\sim 300 \mu m$. The RPDF is also a useful tool to estimate the correlation void ($l$) which is defined as the distance at which probability to find a particle around a reference particle becomes half (i.e $g(r)|_{r=l} =0.5$). It gives an indirect measure of the repulsive force experienced by a dust particle due to other neighboring dust particles. In this particular case, the void length turns out to be $\sim 130 \mu m$. Using the correlation function we have also studied the spatial variation of the inter-particle distance as a function of the radial distance away from the center of the crystal. For doing this, the crystal data has been sampled over small regions of  5 $mm^{2}$ area and the RPDF has been calculated in each region and the inter-particle distance obtained from the position of the first peak. The results for the crystal structure of Fig.~\ref{fig:fig2}.(b) are shown in Fig.~\ref{fig:fig3}.(b). As can be seen the inter particle distance is minimum at the centre and increases as one moves away to the outer edge in a radial direction. This spatial inhomogeneity of a dust crystal is consistent with earlier numerical simulation results of Totsuji et.al \cite{totsuiji} and is also a characteristic feature of two dimensional finite crystalline structures \cite{Filinov2001}. As a further confirmation of the crystalline structure we have also measured the orientation ordering of the dust particle formation. The orientation order of a system is generally quantified by estimating the bond order parameter $(\psi_6)$ \cite{Bonitzbook}. For a perfect crystal, $\mid{\psi_6}\mid$ would be one, but in most experimental situations a value greater than 0.45 is considered to be indicative of a crystalline state \cite{Bonitzbook}. In our experiments, the bond order parameter, corresponding to Fig.~\ref{fig:fig2}.(a) is estimated by calculating the local order parameter for each particle in the system and then averaging over all particles and comes out to be 0.68.

\subsection{Voronoi Diagram and Delaunay Triangulation}
\begin{figure}[ht]
\includegraphics[scale=0.7]{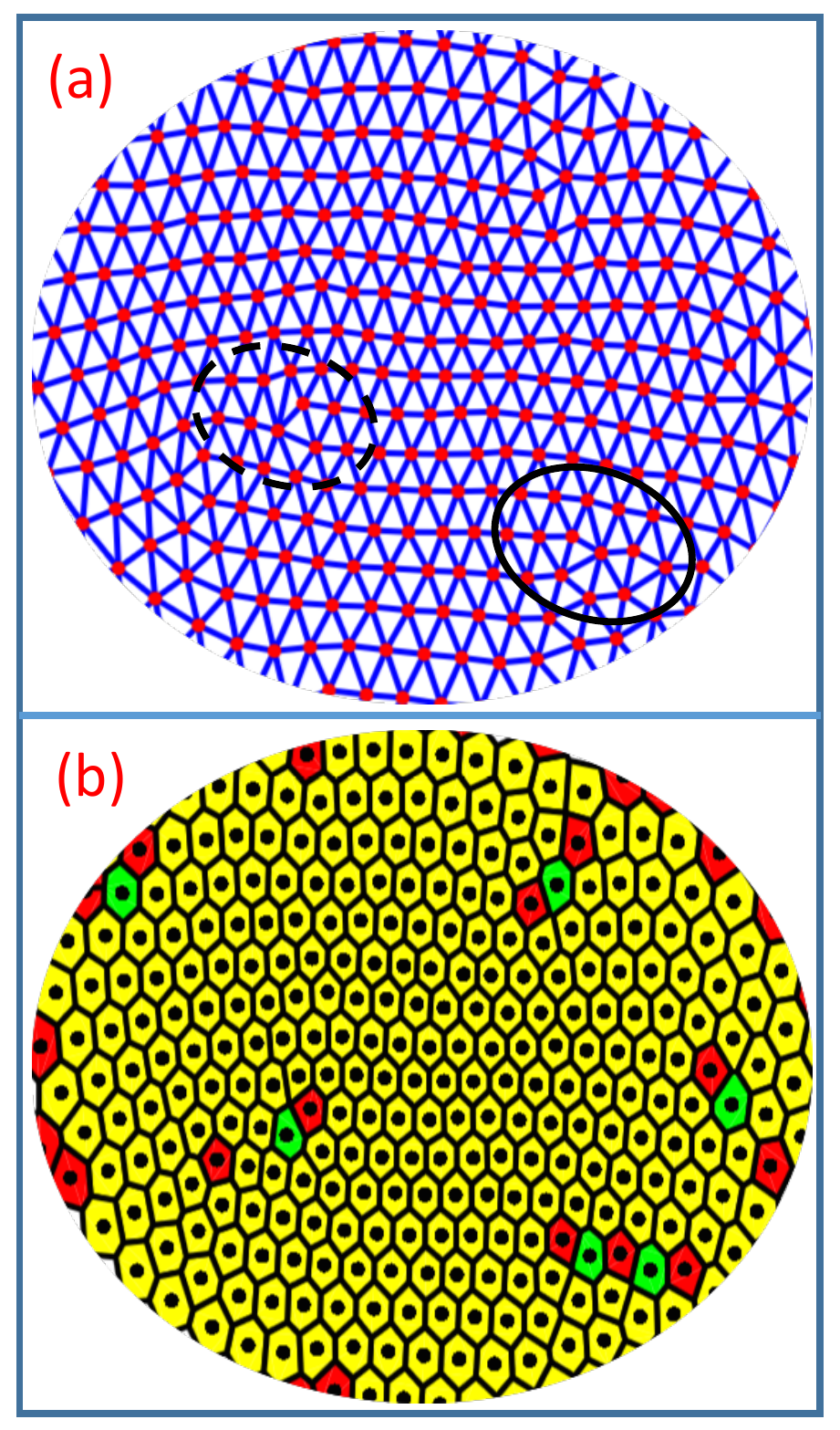}
\caption{\label{fig:fig4} (a) Delaunay Triangulation with defected areas shown in circles  (b) corresponding Voronoi Diagram. }
\end{figure}

 The Voronoi diagram \cite{voronoi} is another useful tool to portray the amount of order (or measure the amount of disorder) in a particular configuration. In the case of the two dimensional dusty plasma crystal it consists of a partitioning of the plane into regions based on distances to each dust position. For each dust particle position there is a corresponding region consisting of all points closer to it than to any other dust particle position. A convenient way to construct the Voronoi diagram is through the Delaunay Triangulation which forms a network of triangles by connecting each dust particle to its nearest neighbours. The perpendicular bisectors of each of these connecting lines (bonds) then creates the Voronoi diagram.  Thus if a particle has six neighboring particles it will be surrounded by a six sided polygon. If there are fewer or more neighbors then it will have polygons of that many sides in the Voronoi diagram and these are classified as defects. Fig.~\ref{fig:fig4}.(a) and (b) show the Delaunay Triangulation and the Voronoi diagram, respectively, for the dusty plasma crystal shown in  Fig.~\ref{fig:fig2}.(b). The yellow colored cells represent hexagonal cells while the other polygons shown in different colors represent defects. 
As can be seen most of the cells are hexagonal indicating a highly ordered crystalline structure with very few defects. In the Delaunay diagram the same information is provided by the number of lines passing through each node  of the diagram and any deviation from six would indicate a defect.
The nature of the defects are highlighted by the two circles shown in Fig.~4(a). The dashed circle shows a disclination in the crystal since in the region inside the circle, seven lines are passing through the node instead of six. Correspondingly in the Voronoi diagram the disclinations appear as green and red polygons that are not hexagons.  An array of disclinations are also visible at the right bottom side of  Fig.~\ref{fig:fig4}.(b). This array of disclinations leads to line defects in the crystal and that can be traced out from Delaunay Triangulation. In  Fig.~\ref{fig:fig4}.(a), the area under the solid circle has line defects in the form of one line splitting into two and bending of grain boundaries. Such defects have also been observed previously in dust crystals created in RF plasma discharges \cite{firstcrystal,Quinn1996}. 
 
 Further, in order to quantify the ordering of the dust cluster a structural order parameter P can be defined, which is the ratio of number of hexagonal structures to the total number of polygons in the Voronoi diagram, 
\begin{eqnarray}
P=\frac{N_H}{N_T}*100 \%
\end{eqnarray}
where $N_H$ and $N_T$ are the number of hexagonal structures and the total number of polygons respectively. The structural order parameter corresponding to  Fig.~\ref{fig:fig2}.(b) is estimated to be 88\%. \\
 \subsection{Estimation of dust temperature and Coulomb coupling parameter using Langevein Dynamics}
\begin{figure}[ht]
\includegraphics[scale=0.8]{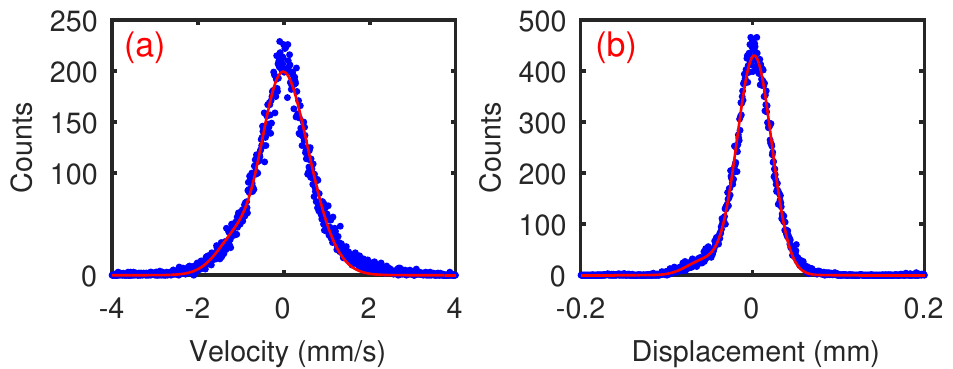}
\caption{\label{fig:fig5} (a) Velocity distribution   (b) Displacement distribution. }
\end{figure}

\begin{figure}
\begin{center}
\includegraphics[scale=0.9]{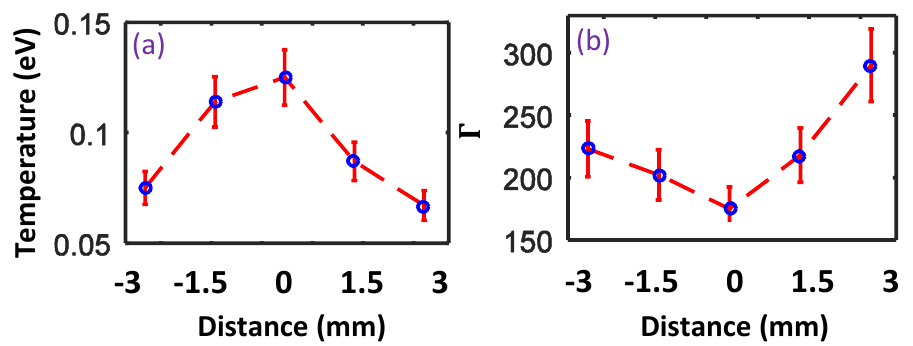}
\end{center}
\caption{\label{fig:fig6}  (a) Dust Temperature and (b) Coupling parameter in radial direction. }
\label{}
\end{figure}

 Dust temperature and the Coulomb coupling constant are two important parameters that help in determining the state  of the dusty plasma system. To find out these parameters, the dust particles are assumed to be distinguishable classical particles and to obey a Maxwell-Boltzmann distribution. In local equilibrium, when the interaction of these dust particles with the plasma as well as with individual neutral atoms is, on average, balanced by neutral friction, the dynamics of individual particles in the lattice can, in principle, be described by a Langevin equation \cite{langevindynamics,morfill2009complex}. In such cases, for each lattice cell one can write the probability distribution \cite{langevindynamics} as, \par
\begin{eqnarray}
P(r,v)\propto exp\left[-\frac{m{(v-<v>)}^2}{2T}-\frac{m{\Omega_E}^2r^2}{2T}\right],
\end{eqnarray}
with all $v$ available in phase space and with $T$ being the particle temperature, $\Omega_E$ the Einstein frequency and $m$ the mass of the dust particle.  The standard deviation of the velocity distribution and of the displacement distribution independently yield the dust temperature and the coupling parameter, respectively. The displacement distribution primarily provides the Einstein frequency which can be used to calculate the coupling parameter once the inter-particle distance is known. The standard deviation of the velocity distribution is given by ${\sigma_v}= \sqrt{\frac{T}{m}}$ and the standard deviation of the displacement distribution is given by ${\sigma_r}= \sqrt{\frac{T}{m\Omega_E^2}}=\sqrt{\frac{\Delta^2}{\Gamma_{eff}}}$, where ${\Gamma_{eff}}=f^2(k)\times\Gamma$ and $\Gamma=\frac{Q^2}{T\Delta}$. $f^{2}(k)$ is the correction factor for Yukawa screening with $k$ as the ratio of inter-particle distance to the Debye length, and for a 2D dusty structure it can be expressed as $3e^{-k}(1+k+k^{2})$. $Q$ and $\Delta$ are the charge acquired by the single dust particle and the inter-particle distance, respectively. Here, $\Gamma_{eff}$ is the modified coupling parameter that takes into account the Yukawa screening potential and its corrections \cite{langevindynamics}.  Fig.~\ref{fig:fig5}. shows the velocity distribution and displacement distribution respectively for our experimental data where approximately 100 particles have been considered for 200 consecutive frames. The dust temperature is estimated to be 0.1 eV from the velocity distribution function and the coupling parameter is found to be about 350 from the displacement distribution function. Hence, it can be concluded that the dusty plasma system is in the crystalline phase as the Coulomb coupling parameter value is much above the theoretical critical value \cite{ikezi1986} of 172. By scanning across the radial width of the crystal we also determine the local variation of the temperature $T$ and the coupling parameter $\Gamma$ as a function of the distance from the center. These variations are shown in  Fig.~\ref{fig:fig6}. The dust temperature is found to be maximum at the center and reduces towards the edge.  At the center, the dust temperature is around 0.12 eV and reduces to 0.08 eV at the edge.  The maxima of the temperature at the centre arises from the strong inter particle potential energy and the high Coulomb pressure at the centre as the inter particle distance is minimum there. A small displacement of the particles from their equilibrium position makes them move more randomly around the equilibrium position. As a result the screened Coulomb potential energy gets converted into kinetic energy which leads to a gain in the temperature at the centre. As a consequence, the coupling parameter shows a  variation similar to the variation of the inter particle distance as shown in  Fig.~\ref{fig:fig3} (b). The coupling parameter has a minimum of 180 at the centre and a maximum at the edge. The high temperature leads to less coupling parameter and less temperature region is found to be of higher coupling parameter even though the inter-particle distance is decreasing at the center. Our present set of analysis suggests that the dust crystal obtained in our experiments is spatially inhomogeneous - a characteristic feature of finite sized two dimensional ordered structures \cite{Filinov2001}. 

\subsection{Characterization of dust crystal with neutral gas pressure}
\begin{figure}
\begin{center}
\includegraphics[scale=0.7]{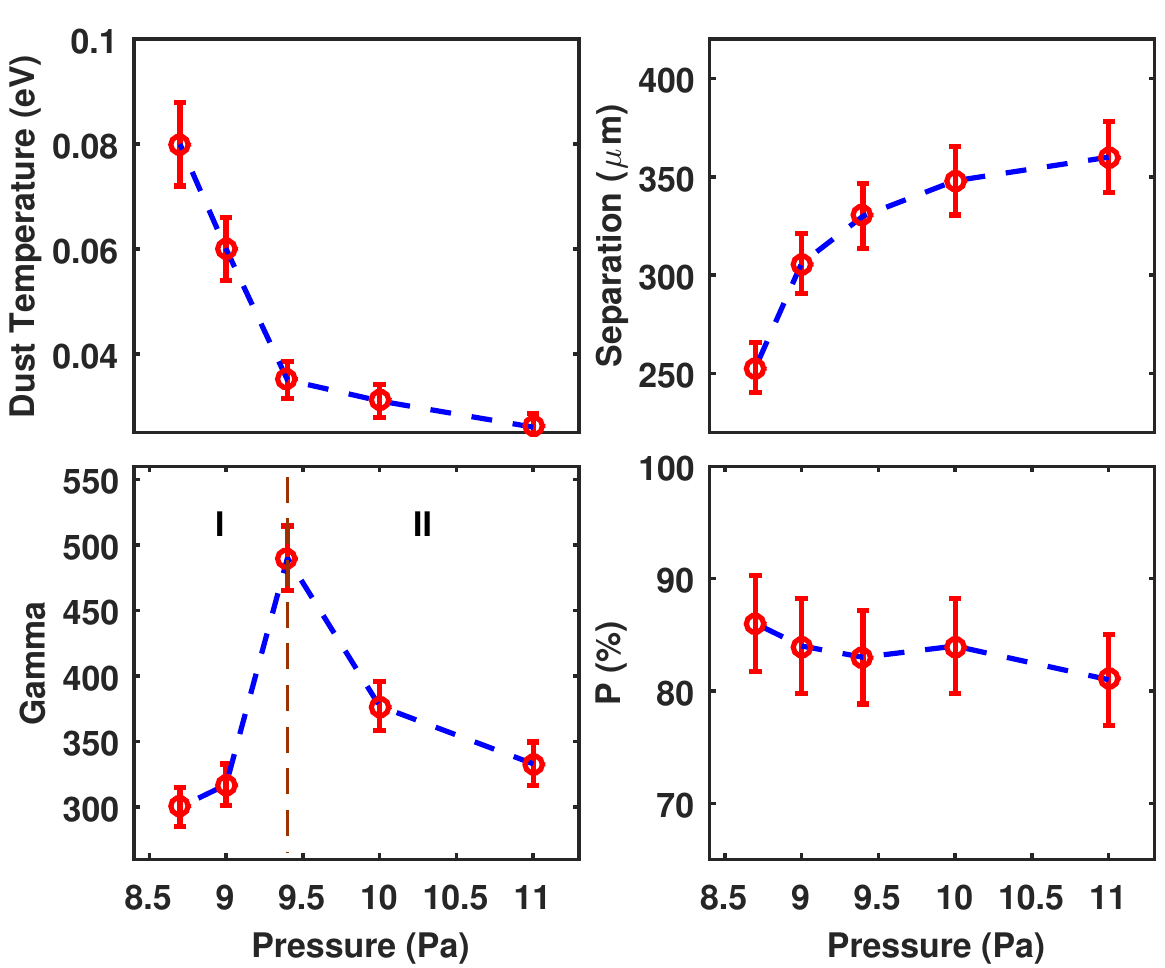}
\end{center}
\caption{\label{fig:fig7} Variation of (a) Inter-particle separation, (b) Dust Temperature (c)  Coupling parameter and (d) structural order parameter(P) with neutral gas pressure. }
\label{}
\end{figure}

We have next studied the dependence of some of the characteristic features of the crystal on the neutral gas pressure in the DC device.  Fig.~\ref{fig:fig7}(a) shows the variation of the inter-particle distance in the central region of the crystal over a range of neutral pressures. It is seen that the inter-particle distance  increases with an increase in the neutral pressure. This can be understood in terms of earlier observations made in the DPEx device on the variation of plasma parameters with pressure \cite{jaiswal2015dusty}.  It has been shown that an increase in the neutral pressure raises the plasma density (due to increased ionization) and lowers the electron temperature (due to friction) which results in a net decrease in the plasma Debye length. This leads to a shrinking of the sheath width and thereby an expansion of the crystal area and hence an increase in the inter-particle distance. The  dust temperature also falls with an increase in the neutral pressure as shown in Fig.~\ref{fig:fig7}(b) which can be directly attributed to the collisional cooling of the dust particles. The dependencies of the inter-particle distance and the dust temperature on the neutral pressure in turn affect the coupling parameter $\Gamma$ whose variation is shown in Fig.~\ref{fig:fig7}(c). The coupling parameter is found to increase with neutral pressure till $p=9.5$ Pa and then it is observed to decrease. This is because, at first even though the inter-particle distance is increasing, the fall in dust temperature is fast which causes the coupling parameter to increase. When the neutral pressure is increased further, the inter-particle distance increases and the dust temperature decreases. However, the fall of dust temperature is no longer so steep. This leads to a decrease in the coupling parameter since the inter particle distance has an exponential inverse dependence on the coupling parameter.  Hence the variation of the coupling parameter with neutral gas pressure as shown in Fig.~\ref{fig:fig7}(c) is divided into two regions. In Region-I, the coupling parameter is temperature dominated whereas in Region-II it is inter-particle distance dominated. It is also to be noted that the coupling parameter remains in the solid state regime over this entire range of pressures. Fig.~\ref{fig:fig7}(d) shows the structural order parameter variation. It is found to be almost constant ($\sim 85\%$) with the change of neutral gas pressure in the range of 8 pa to 11 Pa, which essentially indicates that the order of the dusty plasma crystal does not depend on the neutral gas pressure in this specific range.  
\subsection{Dependence of the dust temperature on the particle size}

To investigate the dependence of dust size on the temperature of dust particles comparisons were made between the characteristics of the two crystalline structures of Fig.~2 under similar discharge conditions.  It was found that the dust temperature of the larger particle crystal was higher at around 2 eV compared to  0.1 eV in case of particles of diameter 4.38 $\mu$m.  According to the capacitive model\cite{dustcharge1,dustcharge2} for spherical capacitor, the charge residing on a dust particle of radius \lq$a$' is estimated to be $Q=CV_s$, where $C=4\pi\epsilon_0a$ is the capacitance of a spherical dust particle, $V_s$ is the surface potential, which can be estimated using a Collision-Enhanced Collections (CEC) model \cite{khrapak2005,khrapak2006}. It essentially indicates that the bigger particle acquires a higher charge than that of smaller particles for a given discharge condition and as a result the bigger particle also suffers higher dust charge fluctuations. This observation agrees well with the theoretical prediction of Vaulina \textit{et al.}  \cite{chargefluctuationheating} in which it is mentioned that the dust charge fluctuation is one of the main mechanisms to heat up the dust particles. Their study also concludes that the dust charge fluctuation directly depends on the two basics parameters of dusty plasma namely the charge acquired by the dust particles and their mass. Consequently, the higher charge leads to higher temperature because of the charge fluctuation heating mechanism, which is exactly observed in our experiments. According to the theoretical model \cite{chargefluctuationheating} the temperature rise from the charge fluctuations is given by 
\begin{eqnarray}
T_f=\frac{\mid{Z_d}\mid^3 e^4}{2m_dl^4}+\frac{m_dg^2}{2\mid{Z_d}\mid\xi}
\end{eqnarray}
where $\mid{Z_d}\mid, m_d, e,g$ are respectively the charge acquired by the dust, the dust mass, the electronic charge and the acceleration of gravity, respectively. $l$ and $\xi$ are parameters that depend on the system dimension, the electron temperature, the ion mass and the dust radius. In order to get a theoretical and experimental comparison of the temperature rise from the charge fluctuation of the dust particles an experiment is carried out to measure the dust temperature at a neutral pressure of 10 Pa, electron temperature of  3 eV and ion density of $10^{15}m^{-3}$ \cite{jaiswal2015dusty} for dust particles of diameters 4.33 $\mu$m and 10.66 $\mu$m. Theoretically, it is found that the temperature rise from the charge fluctuation is $\sim$35 times higher for the case of particles of diameter 10.66$\mu$m compared to the particles of diameter 4.33 $\mu$m. Furthermore, in experimental observations, this value comes out to be $\sim$20. The difference may be arising from some factors of dust heating other than the charge fluctuation mechanism.
\begin{figure}[ht]
\includegraphics[scale=0.6]{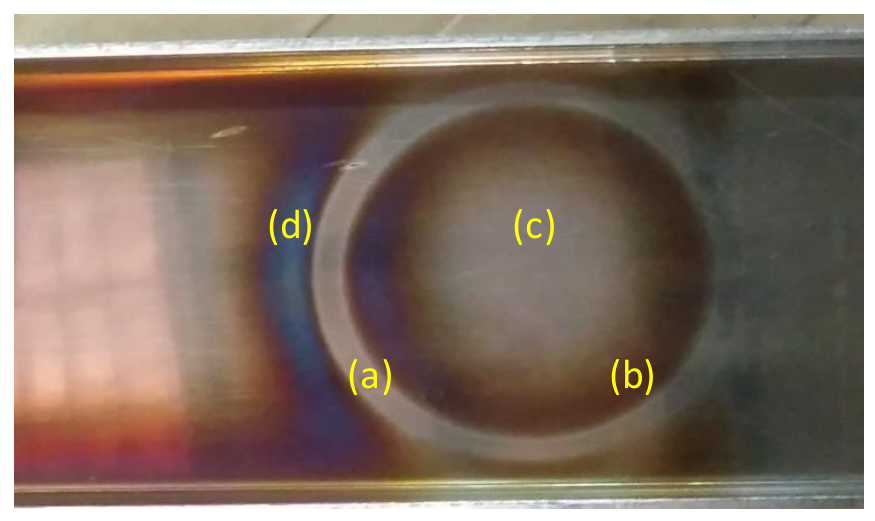}
\caption{\label{fig:fig8} Cathode after exposed to plasma for hours. }
\end{figure}

\section{Summary and Conclusions}
\label{sec:Conclusion} 

To summarize, we have successfully demonstrated the creation and sustenance of dusty plasma crystals in a DC glow discharge plasma over a wide range of discharge parameters. These crystals are long lived and can be sustained for hours by maintaining the background plasma conditions. To establish the crystalline nature of these structures we have carried out a number of diagnostic tests including determination of the radial pair correlation function,orientational bond ordering parameter, Delauney triangulation and Voronoi diagrams of the structures. Other characteristic features of these finite sized crystals such as inhomogeneity in particle spacing and spatial variations of the dust temperature have also been experimentally demonstrated. We have also studied the dependence of the crystal properties on the background neutral pressure and size of the dust particles and delineated these dependencies in terms of the underlying changes in the basic plasma properties sustaining the crystal. \\

On the question of what factors have been responsible for the present success of creation of a dust crystal in a DC glow discharge, we believe our experiments carried out in the DPEx setup  have benefited from two novel features of the device. The asymmetrical electrode configuration (small circular anode and large rectangular cathode) that do not face each other but are configured in the geometry shown in Fig.~\ref{fig:fig1} has helped in reducing the heating effects associated with ion streaming. The ion path has been further influenced by the metal confinement ring which appears to have kept the streaming ions away from the inner central region of the ring where the crystals are formed. We have found experimental evidence of such a behaviour of the ions by examining the surface features of the cathode after several hours of experimental shots and exposure of the cathode to the plasma. Fig.~\ref{fig:fig8} shows a snapshot of the cathode after such an exposure. The regions of strong sputtering arising from the impact of the energetic ions are seen as dark burnt areas. We notice that the ion bombardment is primarily restricted to two circular regions marked as (b) and (d) that encircle the confining ring on the inside and outside. The regions marked (a) and (c) are relatively free of energetic ion impacts - with (a) being just above the ring and (c) being the central region where the dust particles are levitated to form a mono-layer. When the ring is removed we are unable to obtain a proper crystalline structure. Thus the combination of the geometry of the electrodes and the presence of the confining ring play an important role in facilitating the creation of dust plasma crystals in a DC glow discharge plasma. Our findings could be useful for exploring other innovative modifications in various DC glow discharge devices to make them suitable for the study of dusty plasma crystals.\\

\bibliography{reference}
\end{document}